\documentclass[]{aa}

\def\sqr#1#2{{\vcenter{\vbox{\hrule height.#2pt
    \hbox{\vrule width.#2pt height#1pt \kern#1pt \vrule width.#2pt}
    \hrule height.#2pt}}}}
\def\square{\mathchoice\sqr44\sqr44\sqr{2.1}3\sqr{1.5}3}

\usepackage{graphics}

\begin{document}
\nonstopmode

\thesaurus{11(%
11.06.2; % Galxs: fund.pars.
11.09.1; % Galxs: indiv.
11.09.2; % Galxs: IGM
11.16.2; % Galxs: pec.
11.19.6) % Galxs: struc.
	  }

\title{BVRI-H$\alpha$ surface photometry of peculiar ring
galaxies\thanks{Based on observations made at the Observat\'orio do Pico
dos Dias (OPD), operated by the MCT/CNPq/Laborat\'orio Nacional de
Astrof\'{\i}sica, in the state of Minas Gerais, Brazil}}

\subtitle{I. \object{HRG\,2302}}

\titlerunning{Photometry and Spectroscopy of pRGs. I}

\author{M.L.M. Myrrha\inst{1} \and L.P.R. Vaz\inst{1}
\and M. Fa\'undez-Abans\inst{2} \and M. de Oliveira-Abans\inst{2}
\and D.S.L. Soares \inst{1}}

\offprints{M.L.M. Myrrha}

\institute{{Departamento de F\'{\i}sica, ICEX--UFMG, C.P. 702,
30.123-970 Belo Horizonte, MG, Brazil
mfaundez@lna.br, mabans@lna.br
\and
CNPq/Laborat\'orio Nacional de Astrof\'{\i}sica, C.P. 21, 37.500-000,
Itajub\'a, MG, Brazil \\
e-mails: leticia@fisica.ufmg.br, lpv@fisica.ufmg.br, mfaundez@lna.br,
mabans@lna.br, dsoares@fisica.ufmg.br}}

\date{Received $<$date$>$; accepted $<$date$>$}

\maketitle

%%%%% Abstract  %%%%%%%%%%%%%%%%%%%%%%%%%%%%%%%%%%%%%%%%%%%%%%%%%%
\begin{abstract}

Detailed BVRI-H$\alpha$ photometry has been obtained for the ring galaxy
\object{HRG 2302}. This is probably a still interacting system which
shows two principal components: the target, which is a knotted
ringed-disk, and the intruder, an elongated galaxy with two
substructures probably caused by the interaction. The existence of
\ion{H}{II} emission regions is suggested by H$\alpha$ images and
medium dispersion spectra of the three brightest parts of this system.
Analysis of a color-color diagram suggests recent star formation, in
agreement with the behavior of interacting galaxies with nuclear
emission.  \object{HRG\,2302} has been previously classified as a Polar
Ring galaxy (Fa\'undez-Abans\,\&\,de Oliveira-Abans\,\cite{fa8a}), but
we have reclassified it as Elliptical-Knotted, based on the
morphological substructures found.
This work reveals, for the first time, the existence of at least 15
fainter galaxies in the field within $\sim4\arcmin$ around
\object{HRG\,2302}, giving positions and integrated standard magnitudes
for all of them.


\end{abstract}

\input epsf

%%%%% Section 1 %%%%%%%%%%%%%%%%%%%%%%%%%%%%%%%%%%%%%%%%%%%%%%%%%%
\section{Introduction}
\label{intro}

Among the peculiar galaxies, there are objects which take on the
appearance of a smoky ring. These beautiful objects, generically called
Ring Galaxies, can be classified into Normal Ring Galaxies (NRG) and
{\em peculiar Ring Galaxies} (pRG\footnote{Note that PRG (all uppercase
letters) is normally used to denominate Polar Ring Galaxies, as, e.g.,
in Whitmore et al. \cite{w90}, Reshetnikov et al. \cite{r94},
\cite{r95}, \cite{r96}, Bekki \cite{b7}, \cite{b8} and others. Many
indications suggest that most PRGs are pRGs}).  Among the NRG we find
objects with morphology SB and Sa and no evidence of interaction with
other objects. They are formed through resonances in stellar orbits
and dust and gas flows, due to the presence of a bar or an oval
distortion in the galactic gravitational potential (e.g., Buta
\cite{but} and references therein).  The pRG classification, in turn,
is used for objects with distorted structures, off-center nuclei, and
other aspects that may be the result of events such as collision,
merging and tidal interaction. A description of the morphological
characteristics of the pRGs is found in Fa\'undez-Abans \& de
Oliveira-Abans (\cite{fa8a}, hereafter FAOA). Another point in studying
pRGs is that they are useful tools in the study of star formation in
interacting systems. Models indicate that the ring forms when another
galaxy, the intruder, collides almost head-on with a larger rotating
disk galaxy, the target (Lynds \& Toomre \cite{lt}, Theys \& Spiegel
\cite{ts7}, Bekki \cite{b7}, \cite{b8}).  It must be pointed out that
the intruder can be part of the ring galaxy itself, as is the case in
PRGs, or it can be in the close neighborhood.

Studies in the last decade have shown that the gravitational
interaction is the most important factor in galactic evolution and
affects directly properties such as size, morphological type,
luminosity, star formation rate, and mass distribution. The
perturbation in the gravitational potential of the target galaxy
produces radially expanding density waves, which propagate outward from
the nucleus and trigger the birth of stars. Thus, star formation
history is preserved in the radial color distribution of the material
between the rings. Discussions on general properties of Ring Galaxies
can be found in Theys \& Spiegel (\cite{ts6}, \cite{ts7}), Lynds \&
Toomre (\cite{lt}), Toomre (\cite{t1}), Appleton \&
Struck-Marcell (\cite{as}); for a review, see Dennefeld \& Materne
(\cite{dm}).
\begin{table}
\setlength{\tabcolsep}{0.45\tabcolsep}
\caption[ ]{Data for known sources within $10^{\prime}$ around \object{HRG
2302}: coordinates (J2000), distance in arc minutes from \object{HRG
2302}, the logarithm of the flux (mJ) at 4.85 GHz, and the extinction in
the B band, according to NED.}
\label{dat}
\begin{flushleft}
\begin{tabular}{lllllll}
\hline\noalign{\smallskip}
source&${\alpha_{2000}}\atop{\rm ~h~~m~~~~s}$&
${\delta_{2000}}\atop{~~\circ~~~\prime~~~\prime\prime}$&
${\rm dist}\atop{\prime}$&
${\log F}\atop{({\rm mJ})}$& ext \\
\noalign{\smallskip\hrule\smallskip}
\scriptsize{PMN J1634-8058} &\scriptsize{ 16 34 48.79}&
\scriptsize{$-$81 58 19.6}&\scriptsize{9.0}&
\scriptsize{1.82}&\scriptsize{0.47}\\ [-2pt]
\scriptsize{PMNM 162428.3-810826} &\scriptsize{ 16 33 39.98}&
\scriptsize{$-$81 14 53.6}&\scriptsize{9.6}&\scriptsize{1.41}&
\scriptsize{0.46}\\
\noalign{\smallskip\hrule}
\end{tabular}
\end{flushleft}
\end{table}

This is the first of a series of papers on detailed analysis of a
sample of pRGs, based on new photometric (BVRI-H$\alpha$) and medium
dispersion spectroscopic measurements of the ring galaxy components and
their nearest neighbors.  The pRGs selected here are studied as
detailed as possible with medium ($\la$1.5\,m class) size telescopes.
At the same time we are also searching for adequate candidates for
further studies.  Among our aims are the clear identification of the
target and intruder components of the system, their morphological
classification, and to infer the system's dust and gas content.
Such an investigation is important for providing constraints and input
parameters for numerical simulations and analytical modeling of pRGs
(Bekki \cite{b8}, Athanassoula et al. \cite{apb}).

%%%%% Sub-Section 1.1 %%%%%%%%%%%%%%%%%%%%%%%%%%%%%%%%%%%%%%%%%%%%%%%%%%
\subsection{Early Data on \object{HRG\,2302}}
\label{earlydata}

\object{HRG\,2302} has been selected from the recent classification
work by FAOA, which is concerned with the morphology of pRGs. HRG
stands for Hertling Ring Galaxy, using the nomenclature of
Fa\'undez-Abans et al.  \cite{fahr}). The digits are composed by the
SRC/ESO plate number multiplied by 100 plus the ring galaxy number in
order of appearance on that plate for increasing right ascensions.

The morphological classification of \object{HRG 2302} according to the
NASA/IPAC Extragalactic Database (NED) is elliptical, while FAOA
classify it as a PRG. The extinction in the B band is 0.47\,mag (NED)
and the galaxy is located near the Galactic plane ($l = 310\fdg 81$ and
$b = -21\fdg 84$).

\begin{figure}
\parbox[]{0.1cm}{\epsfxsize=8.6cm\epsfbox{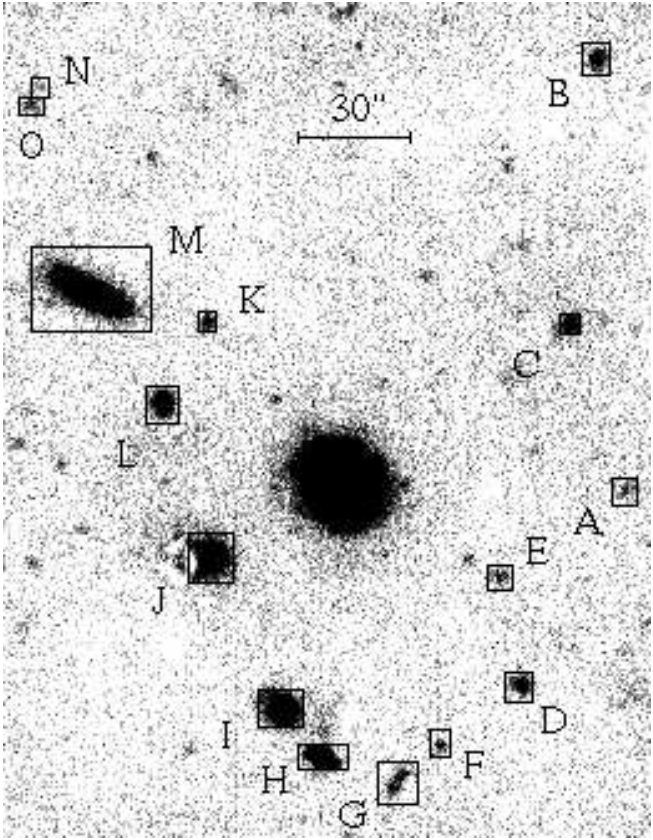} }
\caption{The field around \object{HRG\,2302} in the R filter
(logarithmic scale). Using {\tt IRAF}
standard tools all identified stars have been removed with the
frame's most representative point spread function
(PSF).  In this and in Figs.\,\ref{hpR} to \ref{mdria} North is up and
East to the left.  The ``doughnut'' to the left of object J is
caused by dust particles near the focal plane of the instrument (see
e.g. Gullixson \cite{glx}).
It is not present on
the other frames, where this object's profile also deviates from the
stellar one.}
\label{fld}
\end{figure}

Little is known about \object{HRG\,2302} and its neighborhood.
Only the 2 radio sources given in Table~\ref{dat} were found
in a search for other objects
within $10^{\prime}$ around it in the NED database. The images
reveal, however, the existence of 15 probable non-stellar objects
inside a region of $4^{\prime}$-radius around \object{HRG\,2302}, as
shown in Fig.~\ref{fld}. These objects have been identified through the
comparison of their profiles with the image PSF, from which they
deviate systematically in all frames. They were assigned the
letters A to O, in order of increasing right ascension; their
coordinates and integrated standard magnitudes are given in
Table~\ref{coo}. Curiously,
not even the bright cigar-like galaxy on the northeast (``M''), clearly
visible in the ESO Digitalized Sky Survey plates, has an entry in NED,
so far.
\addtolength{\baselineskip}{-3pt}
\begin{table}
\setlength{\tabcolsep}{0.45\tabcolsep}
\caption[ ]{Coordinates of the centroids (J2000, calculated
differentially from the centroid of \object{HRG\,2302}) and integrated
standard BVRI magnitudes for \object{HRG\,2302} and the 15 objects of
Fig.~\ref{fld}.  The estimated errors in the determination of the
centroid position differences is $<$1 pixel ($\la$$0\farcs 3$ at our
plate scale). However, we adopt a conservative estimate of
$\pm$$1\arcsec$ ($\pm$$0\fs$1) for the error in the positions of the
objects from A to M. The coordinates of HRG\,2302 were taken from NED.}
\label{coo}
\begin{flushleft}
\begin{tabular}{lllrrrr}
\hline\noalign{\smallskip}
\scriptsize{object}&${\alpha_{2000}}\atop{\rm ~~h~~m~~~~s}$&
${\delta_{2000}}\atop{~~~\circ~~~\prime~~~\prime\prime}$&
\multicolumn{1}{c}{B}&\multicolumn{1}{c}{V}&\multicolumn{1}{c}{R}&\multicolumn{1}{c}{I}\\
\noalign{\smallskip\hrule\smallskip}
2302&{ 16 32 35.69}&{$-$81 05 39.7}&  $16.41$&$15.685$&$15.201$& $14.56$\\[-3pt]
    &              &               & $\pm 20$&$\pm 92$&$\pm 76$&$\pm 22$\\
  A &{ 16 32 05.3} &{$-$81 05 41}  &  $22.07$& $22.46$& $20.78$& $19.82$\\[-3pt]
    &              &               & $\pm 96$&$\pm 72$&$\pm 21$&$\pm 37$\\
  B &{ 16 32 08.5} &{$-$81 03 53}  &  $20.30$& $20.00$&$19.513$& $18.63$\\[-3pt]
    &              &               & $\pm 25$&$\pm 12$&$\pm 96$&$\pm 25$\\
  C &{ 16 32 11.6} &{$-$81 05 00}  &  $22.43$& $19.97$&$19.379$& $18.40$\\[-2pt]
    &              &               & $\pm 69$&$\pm 12$&$\pm 92$&$\pm 24$\\
  D &{ 16 32 16.8} &{$-$81 06 30}  &  $21.97$& $20.83$& $19.94$& $18.94$\\[-3pt]
    &              &               & $\pm 60$&$\pm 19$&$\pm 12$&$\pm 26$\\
  E &{ 16 32 19.2} &{$-$81 06 03}  &  $20.98$& $21.56$& $20.52$& $20.73$\\[-3pt]
    &              &               & $\pm 40$&$\pm 32$&$\pm 17$&$\pm 93$\\
  F &{ 16 32 25.5} &{$-$81 06 46}  &   $23.6$&  $22.57$&$21.25$& $19.35$\\[-3pt]
    &              &               &$\pm 1.7$&$\pm 47$&$\pm 18$&$\pm 26$\\
  G &{ 16 32 30.1} &{$-$81 06 54}  &  $22.57$& $21.19$& $20.47$& $20.57$\\[-3pt]
    &              &               & $\pm 84$&$\pm 23$&$\pm 15$&$\pm 70$\\
  H &{ 16 32 38.4} &{$-$81 06 48}  &  $19.40$&$18.695$&$18.026$& $17.56$\\[-3pt]
    &              &               & $\pm 20$&$\pm 95$&$\pm 78$&$\pm 23$\\
  I &{ 16 32 43.1} &{$-$81 06 41}  &  $19.74$&$18.296$&$17.590$& $16.80$\\[-3pt]
    &              &               & $\pm 21$&$\pm 95$&$\pm 77$&$\pm 23$\\
  J &{ 16 32 50.2} &{$-$81 05 58}  &  $20.19$&$18.209$&$17.430$& $16.57$\\[-3pt]
    &              &               & $\pm 22$&$\pm 96$&$\pm 77$&$\pm 23$\\
  K &{ 16 32 50.6} &{$-$81 04 59}  &  $20.30$& $21.92$& $20.70$& $18.48$\\[-3pt]
    &              &               & $\pm 34$&$\pm 43$&$\pm 18$&$\pm 26$\\
  L &{ 16 32 55.5} &{$-$81 05 19}  &  $20.46$& $19.46$&$18.850$& $18.26$\\[-3pt]
    &              &               & $\pm 25$&$\pm 10$&$\pm 82$&$\pm 24$\\
  M &{ 16 33 03.2} &{$-$81 04 52}  &  $17.92$&$17.205$&$16.739$& $16.14$\\[-3pt]
    &              &               & $\pm 19$&$\pm 92$&$\pm 76$&$\pm 23$\\
  N &{ 16 33 08.8} &{$-$81 04 01}  &  $23.29$& $22.38$& $21.78$& $20.25$\\[-3pt]
    &              &               & $\pm 67$&$\pm 22$&$\pm 14$&$\pm 27$\\
  O &{ 16 33 09.5} &{$-$81 04 04}  &  $22.99$& $21.46$&$20.930$& $19.89$\\[-3pt]
    &              &               & $\pm 46$&$\pm 13$&$\pm 95$&$\pm 25$\\
\noalign{\smallskip\hrule}
\end{tabular}
\end{flushleft}
\end{table}
\addtolength{\baselineskip}{+3pt}
%

% ............ Figure 1

The photometric analysis presented herein is based on a set of 13 CCD
images taken through standard BVRI-H$\alpha$ filters. The spectroscopic
analysis is based on 2 medium dispersion (CCD) spectra covering 216\,nm
and centered at $\sim$600\,nm.

Section\,\ref{photobs} describes the photometric observations and the
data reduction process. The global photometric properties are presented
in Sect.\,\ref{imenhance}, together with the morphological structure of
the object. Preliminary spectroscopic data and their interpretation
are presented in Sect.\,\ref{prespec}. The results are
discussed in Sect.\,\ref{discuss}, and the conclusions are presented in
Sect.\,\ref{conclu}.  A detailed description of the galaxies in the
field of HRG 2302 will be given elsewhere.

%%%%% Section 2 %%%%%%%%%%%%%%%%%%%%%%%%%%%%%%%%%%%%%%%%%%%%%%%%%%
\section{Photometric Observations and Data Reduction} %2
\label{photobs}

The photometric observations were obtained in August, 1994, on the
1.6\,m telescope of the OPD (LNA/CNPq), Brazil, equipped with direct
imaging camera \#1 (details in {\small\tt
http://www.lna.br/instrum/camara/camara.html}) and an EEV
770$\times$1152 22.5\,$\mu$m square pixels CCD-05-20-0-202 detector
\#048 (thick, front-illuminated and coated for the UV). The readout
noise was 11.55$\,e^-$ and the gain 10.05$\,e^-$/ADU.  The photometric
BVRI bands are defined as in Bessel (\cite{be}) and given in
the URL above, together with the detector quantum efficiency. The
H$\alpha$ interference filter used has the central wavelength at
670.7\,nm and a FWHM of 10.7\,nm.  The plate scale is $0\farcs 30$/pix.

%%%%% SubSection 2.1 %%%%%%%%%%%%%%%%%%%%%%%%%%%%%%%%%%%%%%%%%%%%%%%%%%
\subsection{Data Reduction} %2.1

CCD image processing and data analysis have been done using standard
procedures, {\tt IRAF} and {\tt STSDAS}.
Photometric standard stars from Landolt (\cite{la}) were observed and
the calibration to the standard
BVRI photometric system was performed in the usual way (see e.g.
Reshetnikov et al. \cite{r94}),
the quality of which may be inferred by the data of
Table~\ref{coef}.
The seeing during the observations was $<2\arcsec$.  The
sky brightness, in mag\,arcsec$^{-2}$ for the BVRI filters, was
21.73, 20.53, 19.89 and 18.25, respectively.
\begin{table}
\setlength{\tabcolsep}{0.6\tabcolsep}
\caption[ ]{Standard deviation
($\sigma$) for the calibration to the standard BVRI system for each
night observed in 1994, together with the number of standard stars used
in the process ($N$).}
\label{coef}
\begin{flushleft}
\begin{tabular}{crrrr}
\hline\noalign{\smallskip}
&\multicolumn{1}{c}{Aug 05}&\multicolumn{1}{c}{Aug 06}&\multicolumn{1}{c}{Aug 07}&\multicolumn{1}{c}{Aug 08}\\
\noalign{\smallskip\hrule}
$\sigma_{\rm B}~(N)$ & $0.059~(15)$ & $0.035~(22)$ & $0.027~(18)$ & $0.065~(8)$ \\ [-2pt]
$\sigma_{\rm V}~(N)$ & $0.025~(12)$ & $0.058~(19)$ & $0.037~(18)$ & $0.066~(8)$ \\ [-2pt]
$\sigma_{\rm R}~(N)$ & $0.020~(12)$ & $0.058~(23)$ & $0.042~(18)$ & $0.046~(8)$ \\ [-2pt]
$\sigma_{\rm I}~(N)$ & $0.052~(12)$ & $0.056~(22)$ & $0.059~(18)$ & $0.042~(8)$ \\
\noalign{\hrule}
\end{tabular}
\end{flushleft}
\end{table}

The PSF was determined based  on more than 20 reasonably bright and
isolated stars on each frame. The removal of foreground stars achieved
with the {\tt IRAF} task {\small\tt ALLSTAR} was better than the
results obtained with {\small\tt PEAK/SUBSTAR}. In this
process, at least 15 candidate non-stellar objects, shown in
Fig.~\ref{fld}, have been found.  Their profiles are larger than and
different from the profiles of the stellar objects, including those
with similar magnitude in the same frame.

The sky background has been estimated in two different ways: (1) by
examining the median count in concentric rings of different widths
centered at the galaxy with the external radius extending to the limits
of the frame (more than 5 times the visual radius of \object{HRG\,2302}
from its center), and estimating the lowest intensity level; (2) by
fitting the sky background with a bidimensional third order spline with
1 to 4 pieces in both dimensions, depending on the frame. All the
residual stars (not completely removed by the PSF), blemishes, and the
objects visible in Fig.~\ref{fld} were isolated from the calculations
of the sky fitting surface by use of rectangular masks. The typical
uncertainty of the background level is (0.5\,-1)\%.
We noticed that the longer the wavelength being imaged the
larger the number of spline pieces for a reasonable sky fitting surface.

%&&&&&&&&&&&&&&&& Section 3   &&&&&&&&&&&&&&&&&&&&&&&&&
\section{Image Enhancement}
\label{imenhance}

To extract as much information as possible from the images, we have
applied the techniques of image enhancement by transform processing
described in Fa\'{u}ndez-Abans \& de Oliveira-Abans (\cite{fa8b}), with
emphasis on the R and I frames due to their better S/N ratio.

%&&&&&&&&&&&&&&&& Section 3.1  &&&&&&&&&&&&&&&&&&&&&&&&&
\subsection{Image transforming}
\label{imtrans}

\begin{figure}
\parbox[]{0.1cm} {\epsfxsize=8.8cm \epsfbox{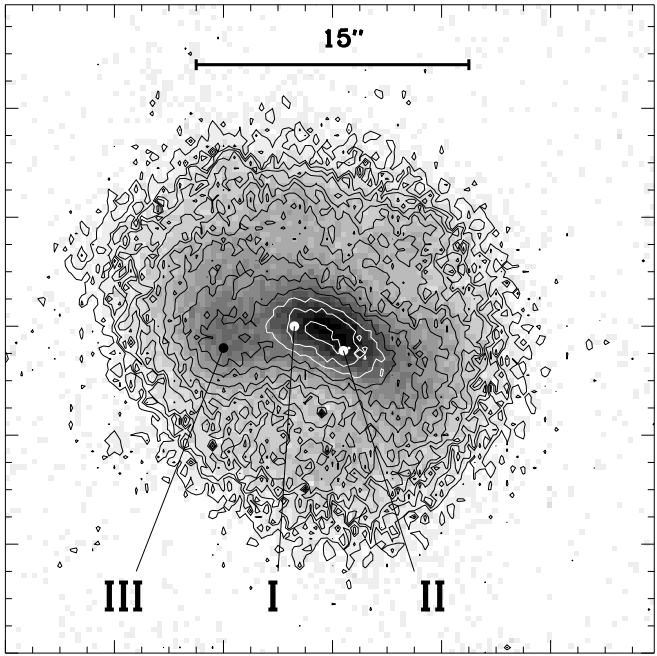} }
%
%LP enhanced.  The field of view in this and in Figs.\,4 to 7 is $54\farcs
%LP 9\times 59\farcs 7$ (183$\times$199 pixels).}
\caption{High-pass filtering of the R image of \object{HRG\,2302}
(intensity scale), with the perturbed isophotes of the central region of
%LP the bulge (R filter, standard magnitudes) overlayed as contour levels.
the bulge (R standard magnitudes) overlaid as contour levels (separated
by $0\fm 2$).
%DS The central bulge and material around the probable clumpy ring have been
The central bulge and material around the probable lumpy ring have been
enhanced by the filtering technique. A filamentary structure connected
to the nuclear region extends along the SW direction. The roman numbers
I and II designate, respectively, the nuclear region and the filament
which form together the bulge. The number III designates the SE bright
ring knot (see Sect.$\,$4).}
\label{hpR}
\end{figure}

In order to enhance high-frequency features, a high-pass filter from
the work cited above has been applied to the R image of
\object{HRG\,2302}.  Five features can be discerned in Fig.\,\ref{hpR}:
a central bulge, an extended faint background nebulosity, an apparently
knotted ring, and two large knots (one on the SE and another on the SW
of the bulge).

% ............ Figure 2

The central region
shows a filamentary structure extending from the nuclear part of the
central bulge in the SW direction, along its major axis. This has been
confirmed in Fourier-hologram experiments.

\begin{figure}
\parbox[]{0.1cm} {\epsfxsize=8.8cm \epsfbox{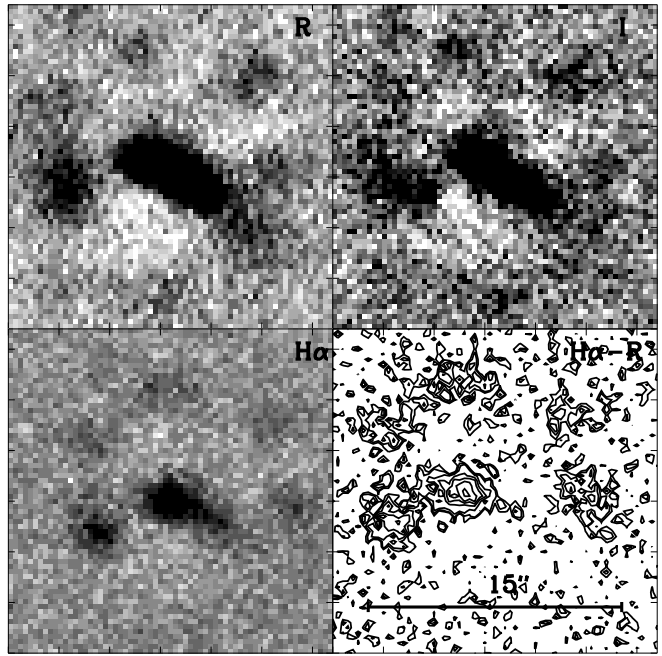} }
\caption{In the upper left, upper right and lower left panels we show
the residual images of the median-filtered frames subtracted
from the original data for the R, I and H$\alpha$ filters, respectively.
In these
panels the lighter areas are artifacts introduced by the method.
In the
lower right panel we show the H$\alpha$ continuum-subtracted isophote
map (H$\alpha-{\rm R}$, intensity scale) of \object{HRG\,2302}.
The isophotal levels are arbitrary.}
\label{mdria}
\end{figure}

The residuals of the subtraction of median-filtered R and I frames
(with a 20$\times$20-pix or $6\farcs 0$$\times$$6\farcs 0$ kernel) from
the original ones (Fig.~\ref{mdria}) highlight the knotty
ring and the elongated nature of the central bulge. The knot III is
bright in both filters and could be composed of two clumps, as opposed
to the other knots (see the panel for the I image in Fig.~\ref{mdria}).

% ............ Figure 4

This technique of subtracting blurred masks has been applied to the
H$\alpha$ image. Median and gaussian filters have been used, both with
kernels of 20$\times$20 pix. The result for the median filter is
displayed in Fig.~\ref{mdria}. The ionized gas is
concentrated in the knots and in regions I and II. They are probable
sites of induced star formation.  Discriminating the filamentary
structure of the feature II is straightforward, as shown in
Fig.~\ref{mdria}.
The structure of \object {HRG\,2302} resembles, in part,
the galaxy NGC 985 --- an object which is more consistent with a tightly
wrapped one-armed spiral plus a linear bar-like structure (see Appleton \&
Marcum \cite{apm}; more on NGC 985 in, e.g., Appleton \& Struck-Marcell
\cite{as87}, Rodr\'{\i}guez Espinosa \& Stanga \cite{res}, Appleton \&
Marston \cite{ama}, and Bransford et al. \cite{b}). On the other hand,
based on this work's data we discern four principal structures in HRG 2302:
a central bulge, an elongated structure extending from the nuclear region
along the SW direction, a prominent knot on the apparent ring, and the
``ring'' itself, composed by several fainter knots.

The filamentary structure mentioned above can also be seen in the
isophotal contour H$\alpha$ map shown in Fig.\,\ref{mdria}. We
subtracted the R image from the H$\alpha$ frame, as an approximation
for the continuum correction. It can be seen that the
features revealed in Fig.\,\ref{mdria} are consistent in all panels.
However, these figures show, also, that
not all nodules or condensations are H$\alpha$ emitters.
%
% ............  Figure 4
\begin{figure*}
\parbox[]{0.0cm}  {\epsfxsize=17.58cm
\epsfbox{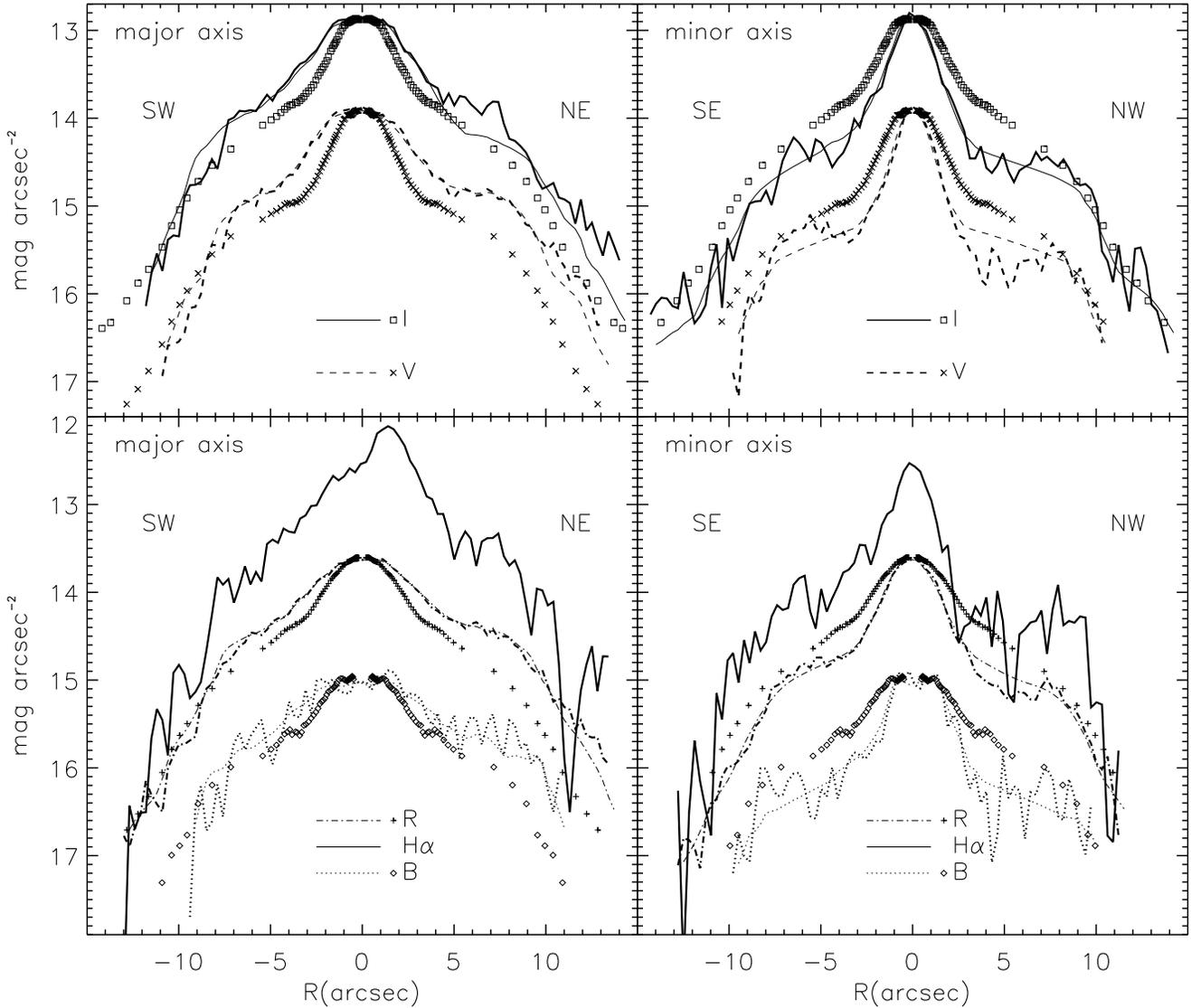} }
\caption{Luminosity profiles in the BVRI passbands (standard magnitudes
arcsec$^{-2}$) and for the H$\alpha$ filter (magnitudes arcsec$^{-2}$
with arbitrary zero point) vs. radius in arcseconds of the observations
(thick lines) and models (thin lines) for \object{HRG\,2302}. The
BVRIH$\alpha$ passbands are represented with different line styles: B,
R, H$\alpha$ (lower panels), V and I (upper panels) along the major
(left panels) and minor (right panels) axes. The equivalent profiles
for the BVRI passbands are also shown, with different symbols for the B
($\diamond$) V ($\times$) R ($+$) I ($\square$) passbands, reflected
around the centroid of the galaxy (see Table 2).}
\label{prof}
\end{figure*}

%&&&&&&&&&&&&&&&& Section 3.2 &&&&&&&&&&&&&&&&&&&&&&&&&
\subsection{Image modeling}
\label{immod}

Isophotal analysis and modeling of \object{HRG\,2302} were performed
through the {\tt STSDAS} tasks {\small\tt ELLIPSE}
and a modified version of {\small\tt BMODEL}, adapted for Ring
Galaxies.

To a greater or lesser extent, all the knots revealed in
this study are present in the B, V, R, and I frames. The
bright knot on the SE (region III) is prominent in all passbands and
there are hints, mainly in the redder images, that it might be
composite, confirming the results of Sect.\,3.1.

%&&&&&&&&&&&&&&&& Section 3.3  &&&&&&&&&&&&&&&&&&&&&&&&&
\subsection{Light and structural parameter profiles}
\label{litestr}

%&&&&&&&&&&&&&&&& Section 3.3.1  &&&&&&&&&&&&&&&&&&&&&&&&&
\subsubsection{Luminosity and color profiles}
\label{lumpro}

Figure \ref{prof} shows the luminosity profiles (LP) of
\object{HRG\,2302} along the mean (both major and minor) axes of the
isophotal ellipses of its central bulge. The isophotal centers and the
directions of the axes obtained with the {\tt STSDAS} task {\small\tt
ELLIPSE} for all the colors were averaged to find the mean major and
minor axes.  The LP for the calibrated BVRI passbands are shown in 3
different ways, while the LP for the H$\alpha$ filter is shown for the
observations only and with an arbitrary zero point.  Firstly, we show
the LP obtained directly from the calibrated images.  Secondly, the
BVRI LP obtained from the model generated in Sect\,\ref{immod} are
traced. These LP are shown as curves with lines of different styles and
thickness. Lastly, we show (as points with different symbols) the
equivalent LP for the BVRI passbands, defined as the elliptical
isophotal magnitude level as a function of the radius of the circle
which has the same area as the ellipse (the equivalent radius, see
de Vaucouleurs \cite{dv48}). The
equivalent LP along the major and minor axes are the same for each
passband and are shown in the panels of
Fig.\,\ref{prof}, folded around the centroid of the galaxy.

% ............  Figure 5
\begin{figure}
\parbox[]{0.1cm}  {\epsfxsize=8.8cm
\epsfbox{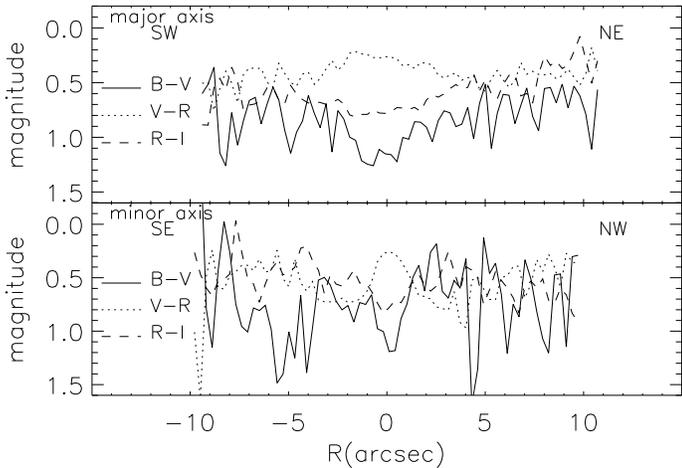} }
\caption{Color profiles (standard magnitudes) B$-$V, V$-$R and R$-$I
vs. radius in arcseconds of \object{HRG\,2302}.}
\label{color3}
\end{figure}

The model fits the observations well in the central part of the galaxy
($-6\arcsec\la R\la 4\arcsec$: major axis;
$-3\arcsec\la R \la 3\arcsec$: minor axis).
The equivalent LP does represent a mean between
the values along the major and the minor axes, and as such the
equivalent LP tends to ignore local deviations due to nodules,
condensations and similar substructures.

Notice that the mean centroid of the BVRI maps, used in derivating the
LP of Fig.\,\ref{prof}, does not coincide with the centroid of the
H$\alpha$ image, which is displaced $\sim$$1\farcs 5$ to the NE. This
can be clearly seen in the H$\alpha$ LP along the semi--major axis,
shown in Fig.\,\ref{prof}. The effect in the centroid is caused by
strong emission from the clump located in the direction of the galaxy
major axis, see Fig.\,\ref{mdria}, and obscuration by dust.

The color profiles of \object{HRG\,2302}, as the calibrated color
indices B$-$V, V$-$R and R$-$I along both mean axes, are shown in
Fig.\,\ref{color3}. The central region with a radius of
$\sim$$5\arcsec$ differs from the rest of the galaxy, showing redder
B$-$V and R$-$I, but slightly bluer V$-$R indices, compared with the
ring.  We identify this central part as being the intruder in this
interaction.

%&&&&&&&&&&&&&&&& Section 3.3.2  &&&&&&&&&&&&&&&&&&&&&&&&&
\subsubsection{Structural parameter profiles}
\label{struct}

The central part, identified with the
intruder, is well fit by ellipses with approximately constant
ellipticity ($\epsilon \sim 0.6$) up to the semi-major axis of around
$9\arcsec$. From this radius, which we identify as the upper limit for
the intruder radius (regions I and II, Sect.\,\ref{imtrans}),
the ellipses become abruptly nearly
circular ($\epsilon \sim 0.1$) in all passbands. In the inner part ($R
\la 3\arcsec$) there is an oscillation in the ellipticity, probably
real, because it is present in all passbands, but possibly strongly
influenced by the seeing conditions. The position angle and the B4
coefficient
(the cos(4$E$) term) follow the pattern, showing
oscillations in this inner region, too. On the other hand, the position
angle does not show the abrupt change at $R\sim 9\arcsec$ and remains
approximately constant ($\approx$60$\degr$), presenting a slight
systematic decrease as the semi-major axis increase.

The B4 coefficients for the B, V, and R images show similar
behavior, not deviating much from zero albeit with predominance of
negative values. This indicates a preference for a ``boxy'' shape in
most of $a \la 10\arcsec$. The behavior in the I
color is different, showing larger negative values.
This is
probably due to the relatively poor quality of the I images, as this
tendence is not evident in the other colors.

% &&&&&&&&& Section 4 &&&&&&&&&&&&&&&&&&&&&&&&&&&&&&&&&
\section{Preliminary spectroscopy}
\label{prespec}

Spectroscopic observations were performed with the 1.6m-telescope at
the OPD in August 26, 1998, equipped with a Cassegrain spectrograph and
CCD {\#}106 (1024$\times$1024 square pixels, 24\,$\mu$m each), with
4.1$e^-$ readout noise and 5.0$\,e^-$/ADU gain.  The grating of 600
lines/mm was centered at 600\,nm (dispersion of 84.1{\AA}/mm and
resolution of 2.0 pixels FWHM). Two spectra (20 min each) were
obtained.  The $3\arcsec$ slit covered the brightest region of the
galaxy. The slit was aligned with the EW direction in order to include
the probable signatures of the intruder (regions I and II, see
Fig.\,\ref{hpR}) and the target galaxy (region III).

% ............  Figure 6
\begin{figure}
\parbox[]{0.1cm}  {\epsfxsize=8.7cm
\epsfbox{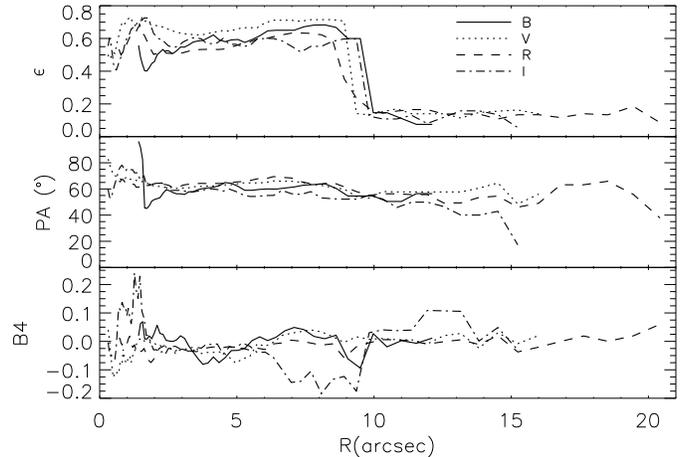} }
\caption{Isophotal parameters of \object{HRG\,2302}: ellipticity,
position angles (in degrees, eastwards from the North)
and the Fourier coefficient of the cos(4$E$) term ($E$ is the
ellipse eccentric anomaly), B4, versus the semi--major axis in
arcseconds. }
\label{epb4}
\end{figure}

The stars used for extinction and flux calibrations are tertiary
standards from Baldwin \& Stone (\cite{bs}), as revised by Hamuy et al.
(\cite{ha2}, see also Hamuy et al. \cite{ha4}). The spectrum reduction
has been done according to standard procedures employing {\tt IRAF}.

\begin{figure}
\parbox[]{16cm} {\epsfxsize=8.8cm \epsfbox{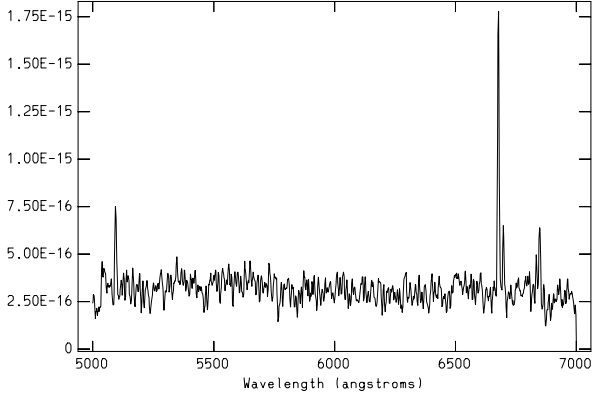} }
\caption{Spectrum of the most prominent part of the intruder, region I,
smoothed by a kernel of 3\,pix. The vertical axis is the calibrated
flux in units of erg\,sec$^{-1}$\,cm$^{-2}$\,{\AA}$^{-1}$.}
\label{spi}
\end{figure}

A velocity of 5\,996$\pm$21 km/s for the region I has been computed
(z~$\sim$~0.020). Within the errors, the spectra of the three
regions are not significantly shifted with respect to each other.
Whether region I suffers a slow braking and this scenario is a result
of a polar collision (see, e.g., Bekki \cite{b8}) remains yet to be
confirmed by better resolution spectroscopic observations, a project now
in progress.
The three
regions exhibit H II-region emission-type spectra with strong
H$\alpha$, [\ion{N}{II}], [\ion{S}{II}] and [\ion{O}{III}] emission
lines (see Fig.\,\ref{spi}).

% ............. Figure 7
%

%spectrum table
\begin{table}
\caption[ ]{Preliminary parameters from the spectrum}
\label{spec}
\begin{tabular}{llll} \hline
\noalign{\smallskip}
\multicolumn{1}{c}{Parameter} & \multicolumn{3}{c}{Regions} \\
\cline{2-4} \\[-8pt]
  & I & II & III \\
\noalign{\smallskip}\hline
$\lambda\lambda$(6716/6731) & 0.96 & 0.43 & 0.75 \\
$N_\mathrm{e}$(${\rm cm}^{-1}$) & 700 & 10,100 & 1,500 \\
$T_\mathrm{e}$(K) & 20,500 & -- & -- \\
\noalign{\smallskip}
\hline
\end{tabular}
\end{table}

The enhanced images of region III show clumps which may be the origin
of most of the emission lines from this region. Whether region III is
the displaced bulge of an original disk galaxy that has been disrupted
due to a collision or whether it is the most prominent \ion{H}{II} region under
formation on the ring, is a matter of further discussion.

An attempt to determine the electron density of the regions I, II, and
III has been made using the [\ion{S}{II}] $\lambda\lambda$ 6761/6731
ratio and the [\ion{N}{II}] relation $\lambda\lambda$ (6548 +
6583/5755), as listed in Table~\ref{spec}. The lack of $\lambda$5755 in
the spectra of both region II and III and of $\lambda$6548 in region II
has not allowed an estimate of the electron temperature in
these regions. As the H$\beta$ line was out of the observed range, a
similar estimation of chemical abundances has not been possible.
Nevertheless, a
preliminary lower limit for oxygen abundance of
12$+$log[O/H]\,$\sim$\,8.3 has been estimated for region I (see also
next section).
Although being  relatively  low, this value
is in agreement with recent oxygen  abundance  determinations  from optical
long-slit spectra of individual  star-forming knots embedded in collisional
rings by Bransford et al. (\cite{b}).

% &&&&&&&&&&&&&&& Section 5 &&&&&&&&&&&&&&&&&&&&&&&&&&&&
\section{Discussion}
\label{discuss}

The images of \object{HRG\,2302} have revealed a knotty ring around an
elongated perturbed bulge. There are two prominent knots: the one
located in region III and another, located to the SW, that has
no significant emission in H$\alpha$. The bulge is composed by two
substructures: a round compact one (region I) and a filament in the
direction of the SW prominent knot (region II).  If this object is a
result of a galactic collision as, for example, in Lynds \& Toomre
(\cite{lt}, Figs. 5 and 6), it seems reasonable to suggest that region
III could be part of the target galaxy's tenuous bulge, which has been
displaced during the encounter (the disk would have given rise to the
ring), and that regions I and II together are part of the perturbed
(disrupted?) projectile galaxy. The H{$\alpha$} image indicates the
presence of bright \ion{H}{II} regions in the central object as well as
in some of the knots. These may be regions of induced star formation.
On the other hand, there seems to be no obvious emission from the SW
knot (see Fig.\,\ref{mdria}).

% ............  Figure 8
\begin{figure}
\parbox[]{16cm}  {\epsfxsize=8.8cm
\epsfbox{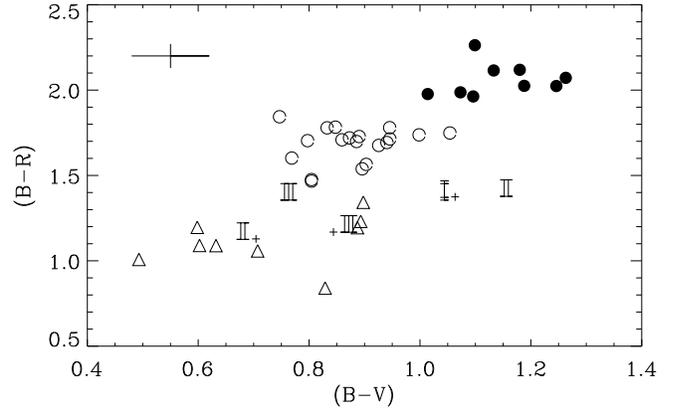} }
\caption{Color-color diagram for \object{HRG\,2302} and
\object{HRG\,54103}.  The filled and the open circles represent the
nucleus and the ring of \object{HRG\,54103}, respectively. The
$\triangle$ symbols represent the knots on \object{HRG\,2302}'s
ring. Regions I, II and III are represented by these same letters; the
ones without subscript represent the region of maximum
intensity, which are the same for all filters; the ones with
the subscript $+$ represent regions of maximum
color index as seen in the color maps.
The 2$\sigma$ ($\sigma$=$0\fm 07$) error bars in each axis are
shown in the upper left part of the panel.}
\label{coc}
\end{figure}

The (B$-$R) versus (B$-$V) diagram in Fig.\,\ref{coc} contains
integrated photometry of selected 4$\times$4-square pixel areas of the
bulges and rings of \object{HRG\,2302} and \object{HRG\,54103}. The
latter (Fa\'undez-Abans et al.\cite{fam8}) has been included for
comparison and because it shows no evidence of interaction
(Fa\'undez-Abans \& de Oliveira-Abans \cite{fa8b}), and has been
cataloged as NRG by Buta (\cite{but}). Thus, a different behavior is to
be expected when compared with an object which has probably undergone
interaction. The average mean error in both colors is less than 0.07
mag.  The measurements of regions I, II, and III are represented in
Fig.\,\ref{coc} by these same numbers. The spot of maximum intensity in
each region (represented with subscript +) does not, in general,
coincide with that of larger color indices, that is, the brighter
portions may not be the redder ones. Thus, two measurements per region
have been made.
No systematics is shown by these measurements except for the fact
that all fall in the same B$-$R$\,<\,1.5$ region occupied
by the ring knots.
The ring is represented by open triangles and nine
measurements have been made along its surface, on the most prominent
clumps. The bulge and ring of \object{HRG\,54103} are represented by
filled circles and open circles, respectively.  An inspection of
Fig.\,\ref{coc} shows that \object{HRG\,2302} is definitely bluer than
\object{HRG\,54103}. This may be caused by localized star formation, a
result confirmed by this work's spectra, which resemble those of
\ion{H}{II} regions. The extreme II data point on the right corresponds
to a spot that does not appear in H$\alpha$ images; this is probably a
neutral substructure superimposed on the blue filament.

The colors of the knots lie in the range $0.4 < ({\rm B}-{\rm V}) <
0.9$, agreeing with early photoelectric measurements by Theys \&
Spiegel (\cite{ts6}).  Evidences from the color profiles
(Fig.\,\ref{color3}) point to classifying the intruder as an early-type
system. Both along the major and minor axes, R$-$I ranges from
$0.5-0.7$, within $5\arcsec$ from the center. These are typical
early-type galaxy colors (see Poulain and Nieto 1994). Although rather
noisy, the B$-$V color ranges from 0.8 to 1.1, also around the typical
value of 1.0 found in elliptical and lenticular galaxies. V$-$R has
overall values ($0.2-0.5$) that are systematically lower than those
measured in early-type galaxies, which is likely the effect of dust
extinction. The presence of dust is expected due to the disruption of
the target galaxy, seemingly a gas-rich late-type system.

% ............  Figure 9
\begin{figure}
\parbox[]{0.1cm}{}
\begin{picture}(108., 186.)(0,0)
\put(40,0) {\epsfxsize=5.8cm \epsfbox{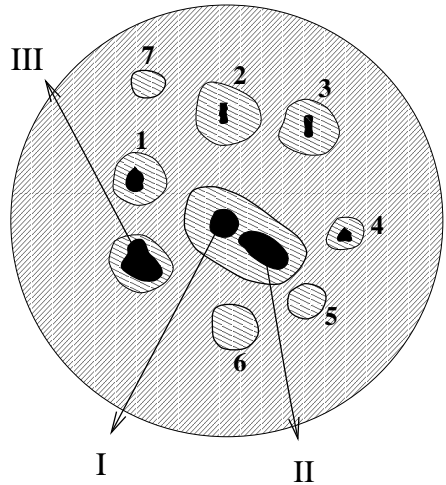} }
\end{picture}
\caption{Simplified sketch of \object{HRG\,2302}. Black regions are
H$\alpha$-emitters
while dashed ones are only visible through broad-band filters. The circular
dashed region represents the extended faint background nebulosity. In the
suggested collisional model for the system, structures
I and II form the intruder  and III is the displaced nucleus of the target
galaxy.}
\label{geom}
\end{figure}

There seems to be a radial color gradient, a known property of
ring galaxies. The regions inside the knots are generally redder both
in (B$-$V) and (B$-$R). These color changes are consistent with the
simple starburst model in which the stars are formed in the ring and
evolve in the wake of the ring causing the redder colors observed there
(cf. Appleton \& Struck-Marcell \cite{as}). The global (B$-$V) color of
the ring is $0.75\pm 0.15$ (1$\sigma$), which is, within the errors,
the same as the median (B$-$V) of $0.52\pm 0.10$ found by Marston \&
Appleton (\cite{ma}) for a sample of twelve northern ring galaxies. In
order to know how large the influence of dust absorption in the radial
color gradients is, one would require near-infrared observations of the
system.

Plotting \object{HRG\,2302} in the line-ratios diagnostics diagram of
forbidden-line strengths for interacting galaxy nuclei, from Keel et al.
(\cite{k85}), places it inside the sector defined by nuclear
\ion{H}{II} regions of interacting galaxies. The line-ratios obtained in
the present work are:
log([\ion{O}{III}]$_{\lambda\,5007}$/H$_{\alpha}$)\,=\,$-$0.78 and
log([\ion{N}{II}]$_{\lambda\,6584}$/H$_{\alpha}$)\,=\,$-$0.64.
Similarly, we estimate for the nuclear and disk H$\alpha$+[\ion{N}{II}]
emission equivalent widths the values: 63\,m{\AA} for the central bulge
(regions I and II) and 69\,m{\AA} for region III. Plotting these numbers
in Kennicutt et al.'s (\cite{k87}) correlation diagram, places this pRG
in the intersection of the areas defined by interacting and
non-interacting galaxies, closer to the interacting one.

Fosbury \& Hawarden (\cite{fh}) estimated elemental abundances for two
of the HII regions on the ring of A0035 (the "Cartwheel" galaxy) and
found low metallicity values: 12 + log[O/H] = 8.10 and 7.96.
This work's
oxygen abundance of at least 8.3 is then in agreement with the
above-mentioned estimates. The existence of induced star-forming in the
perturbed intruder as well as in the target is thus well characterized.

One can estimate the recent global star formation rate (age $\la 10^9$
yr) in HRG 2302 from the blue luminosity (Sandage \cite{san}, Young et
al. \cite{y96}). We get $3.10\times 10^9$ L$_{\odot,{\rm B}}$ from the
data obtained here and H$_\circ=69$ km\,s$^{-1}$\,Mpc$^{-1}$ (Ferrarese
et al. \cite{fmk17}), implying
an overall star formation rate, normalized to unit area, of $1.83\times
10^{-8}$ M$_{\odot}$ yr$^{-1}$ pc$^{-2}$. This is approximately 5 times
as large as the average rate for late-type galaxies (Buat et
al.\,\cite{bdd}, Elmegreen \cite{elm}).

% &&&&&&&&&&&&&&& Section 5.1 &&&&&&&&&&&&&&&&&&&&&&&&&&&&
\subsection{Geometry and composition}
\label{geomcomp}

With the results obtained in the present analysis it is possible
to speculate on the geometry and the overall
structure of \object{HRG\,2302}. The main conclusions are represented
in Fig.\,\ref{geom}.
With the adopted $H_\circ$,
a distance of $86 \pm 9$ Mpc has been
derived. Table \ref{tabdim} displays the various components' estimated major
and minor axes; the associated errors are 1.5 pix, 0.45 arcsec, and 0.19 kpc.

\subsection{Alternative scenarios}
\label{altern}

Two tentative scenarios might be added to the one already discussed in the
text. The final discrimination amongst them will be delayed until detailed
spectroscopic studies as well as full mapping in the near-infrared and radio
bands are performed.

Firstly, the possibility of a nuclear bar is considered.  HRG 2302 is a dwarf
galaxy, with integrated blue magnitude of --18.3 (similar to the LMC,
for example). Regions I and II constitute a slightly inclined bar; region II
has a foreground counterpart behind region I being thereby obscured.
The B4-coefficient profile is consistent with a bar morphology since
it shows negative values in most of the region within $10\arcsec$.
Near-infrared observations are likely to uncover the entire bar
morphology. The system is then a very early SBaV galaxy with the tightly
wound arms resembling a knotted ring. The peak of H$\alpha$ emission in
region I, which is the assumed galaxy nucleus, is consistent with what is
expected from gas being driven to the center, as derived from barred
galaxies modeling (e.g., Roberts et al. \cite{rob}, Sakamoto et al. \cite{sak}).

Secondly, the possibility of a moderately fast encounter between HRG 2302 and
another galaxy is considered. We suggest here a small-scale version
of what may be happening in the Cartwheel system according to the mechanism
put forward by Higdon (\cite{hig}). Based in HI observations of the system,
he attributes to a third neighbor galaxy, named G3 and other than the two
closest to the Cartwheel's disk, the responsibility for generating the
peculiar morphology observed in the system. G3 is shown to be linked
to the main galaxy by a broad HI plume, at least in one channel map,
suggesting an earlier passage ($\approx$ 300 Myr-old) with a moderate
relative velocity of about 200 km s$^{-1}$ (see Higdon's Fig.\,17).
HRG 2302 forms with galaxy M (see our Fig.\,1 and Table 2) a comparable
system. We must point out however that the comparison is based on a number
of circumstantial evidences since there is only one galaxy redshift
measured in the field and one does not really know whether galaxy
M is physically bound to HRG 2302. Their piojected separation
and sizes though are consistent with the assumption. Also, the magnitude
difference between Cartwheel and G3 is about the same as in HRG 2302
and M, $\Delta M_{\rm B}=$ 1.7 and 1.5 respectively. The linear
separation in the least massive pair --- at the redshift of HRG
2302 --- is scaled down with total luminosity as it would
be expected for similar dynamical systems: the separation
is 88 kpc in the
Cartwheel-G3 pair (Higdon \cite{hig}) and 38 kpc in HRG 2302-M. Thus we
have a case where a detailed mapping in the 21-cm line of neutral
hydrogen would be highly desirable (project for a 21-cm line imaging
of the \object{HRG\,2302} region underway).
The role of regions I, II and III should be readdressed here as well.
We identify regions I and II as the disturbed nuclear region of
the target disk galaxy, and region III as a more
massive and active H II cloud along the ring.

% &&&&&&&&&&&&&&& Section 6 &&&&&&&&&&&&&&&&&&&&&&&&&&&&
\section{Conclusions}
\label{conclu}

In this article we present the results of CCD surface photometry and a
preliminary spectroscopy for the system \object{HRG\,2302} which is
studied for the first time in the literature. The accuracy of this
work's photometry is better than 0.06 mag in all passbands.
\begin{table}
\setlength{\tabcolsep}{0.45\tabcolsep}
\caption[ ]{Estimated dimensions (major and minor axes)
of the substructures. ``Outer'' is a
lower limit for the diameter of \object{HRG 2302}, as measured on the radial
intensity profile, at 10\% of the sky level in the R filter.
The knots are shown in Fig.\,\ref{geom}.
}
\label{tabdim}
\begin{flushleft}
\begin{tabular}{lrrrrrccrrrrr}
\hline\noalign{\smallskip}
region~~& \multicolumn{2}{c}{2a} && \multicolumn{2}{c}{2b} &~~~& knot& \multicolumn{2}{c}{2a} && \multicolumn{2}{c}{2b} \\
	  \cline{2-3} \cline {5-6} \cline{9-10} \cline{12-13}\\ [-8pt]
          &$^{\prime\prime}$&   kpc&&$^{\prime\prime}$&   kpc&&
          &$^{\prime\prime}$&   kpc&&$^{\prime\prime}$&   kpc\\
\hline\noalign{\smallskip}
outer   &30.0&12.5&&-- &  --&&1  & 2.1&0.87&& --&  --\\ [-2pt]
ring    &13.2& 5.5&&-- &  --&&2  & 2.1&0.87&& --&  --\\ [-2pt]
I   & 2.7& 1.1&&2.0&0.83&&3  & 3.4& 1.4&&2.2&0.91\\ [-2pt]
II  & 3.0& 1.2&&1.5&0.63&&4  & 2.7& 1.1&& --&  --\\ [-2pt]
I+II& 7.3& 3.0&&3.5& 1.5&&5  & 4.5& 1.9&& --&  --\\ [-2pt]
III & 4.1& 1.7&&3.0& 1.2&&6  & 2.4& 1.0&& --&  --\\ [-2pt]
&&&&&&&7  & 3.2& 1.3&& --&  --\\ [-2pt]
\noalign{\smallskip\hrule}
\end{tabular}
\end{flushleft}
\end{table}

The main results and conclusions of the present work are: (a)
\object{HRG\,2302} is probably an interacting Ring Galaxy, (b) it has
been assigned the type of Elliptical-Knotted, based on FAOA (and as
opposed to their previous classification of Polar Ring), (c) its
spectrum is characteristic of \ion{H}{II} regions, (d) several
substructures have been revealed, which suggest the favored scenario of an
intruder having collided with a face-on disk galaxy, (e) detailed
surface photometry and spectroscopy help disentangle the morphological
details which may lead to a better classification and pRG family
understanding,
(f) two alternative scenarios for the formation of the \object{HRG\,2302}
system are presented,
(g) for the first time, at least 15 non-stellar objects,
all probably (some certainly) galaxies, have been revealed within
$4^{\prime}$ around \object{HRG\,2302}.

A detailed study of the rich field of galaxies around
\object{HRG\,2302} is now in progress.

\begin{acknowledgements}

This work was partially supported by FAPEMIG grant CEX 1864/95, CNPq,
FINEP, and CAPES (Brazilian institutions). This research has made use
of the NASA/IPAC Extragalactic Database (NED, operated by the Jet
Propulsion Laboratory, Caltech, under contract with NASA) and of the
SIMBAD data base (operated by the CDS, Strasbourg, France).
The help of the LNA staff at OPD is gratefully
acknowledged.  We gratefully thank Dr. Bo Reipurth for a critical
reading of the manuscript. We thank an anonymous referee for
relevant comments which improved the presentation of this work.
\end{acknowledgements}

\end{document}